\documentclass{pasj00}

\SetRunningHead{Astronomical Society of Japan}{Usage of \texttt{pasj00.cls}}
\usepackage{times}

\begin{document}

\title{Low-Resolution Spectroscopy of the Recurrent Nova T Pyxidis at its Early Stage of 2011 Outburst}
\SetRunningHead{K. Imamura and K. Tanabe}{Low-Resolution Spectroscopy of the T Pyx 2011 Outburst}

\author{Kazuyoshi \textsc{Imamura}$^1$ and Kenji \textsc{Tanabe}$^2$
        }
\affil{$^1$Department of Mathematical and Environmental System Science, Faculty of Informatics,\\
       Okayama University of Science, 1-1 Ridai-cho, Kita-ku, Okayama 700-0005}
\email{imako@pc.117.cx}
\affil{$^2$Department of Biosphere-Geosphere Science, Faculty of Biosphere-Geosphere Science,\\
       Okayama University of Science, 1-1 Ridai-cho, Kita-ku, Okayama 700-0005}
\email{tanabe@big.ous.ac.jp}

\Received{2012 October 2}
\Accepted{2012 October 17}


\maketitle

\begin{abstract}

  We present our observational results of the recurrent nova T Pyxidis at its early stage of 2011 outburst, 
using a low-resolution spectrograph ($R\approx400$) attached to a 28cm telescope. 
Total nights of our observation are 11, among which 9 nights are during the pre-maximum stage. 
As a result we have obtained a detailed evolutional feature of this recurrent nova 
on the way to its maximum light. 
At first, on the earliest three nights ($-25 \sim -21$ days before maximum), 
broad and prominent emission lines such as 
Balmer series, He I, He II, N II, N III and O I together with P Cygni profile are seen on the spectra. 
The blueshifted absorption minima of H$\alpha$ yields a maximum expansion velocity of 
approximately 2200  km s$^{-1}$, and the velocity gradually decreases. 
Then, Helium and Nitrogen lines are weakened day by day. 
After that (18 days before maximum light), Fe II (multiplets) lines emerge on the spectra. 
These lines are then strengthened day by day, and the P Cygni profiles also become more prominent.  
Accordingly, the expansion velocities turns to be gradual increase. 
In addition, during the pre-maximum stage, 
nova spectral type of T Pyx is thought to evolve from He/N type to Fe II one.

\end{abstract}

\section{Introduction}
   Recurrent novae (RNe, for abbreviation hereafter) are distinguished from the classical novae 
(CNe, hereafter) according to multiple outburst records or single one, respectively. 
However, all of the CNe are thought to have much longer recurrence because of their cataclysmic binary properties. 
By the end of 2011, total definite number of RNe detected is 10 (for known RNe, e.g. see \cite{sch2010}).

  RNe are categorized into three distinct subtypes (Warner 1995, 2008), 
namely T CrB, U Sco and T Pyx subtype.
The criterion of the classification is mainly based on the binary orbital period, 
decline rate (speed class) and type of the secondary star.
Among them T Pyx subtype has long a single member
\footnote{IM Nor is thought to be a possible new member of T Pyx subtype, 
but \citet{kat2002} proposed that IM Nor and CI Aql comprise a new subtype of RNe 
for their massive ejecta and long recurrence times.}, 
i.e., T Pyxidis itself.
The T Pyx subtype are characterized by very short orbital period, and comparatively slow decline. 

   The recurrent nova T Pyx was first detected in 1902 by H. Leavitt (\cite{lea1913}) 
and had then experienced three outbursts 
(before Leavitt, one outburst had been recorded in 1890; \cite{lea1920}).
Next following outbursts were 1920, 1944 and 1966.
Based on these historical records, its recurrence had been thought to be approximately 20 years. 
Nevertheless, we could not have detected any outburst since 1966 
(see VSOLJ \footnote{Variable Star Observers League in Japan (VSOLJ);\\ \tt http://vsolj.cetus-net.org/index.html} 
and AAVSO \footnote{American Association of Variable Star Observers (AAVSO);\\ \tt http://www.aavso.org/} database).

   The above mentioned five outbursts show identical light curves, each with a peak 
visual magnitude of 6.4 and a $t_2 = 32$ days, $t_3 = 62$ days (\cite{sch2010}), 
where $t_2$ and $t_3$ are the amount of time taken for the brightness to decline by 2 and 3 mag 
from maximum, respectively (\cite{pay1957}). 
Although three outbursts had been observed after discovery, rather fragmental spectrum was obtained. 
Among them, \citet{cat1969} had obtained pre-maximum phase spectra for two nights during 1966 outburst.
The orbital period was determined by \citet{uth2010} to be 1.83 hours based on their time-resolved 
optical spectroscopic observations at quiescence. 
They derived its inclination of the binary system to be $i \sim \timeform{10D}$ and 
its white dwarf (WD) mass $M_{\mathrm{WD}} = 0.7 \pm 0.2 \MO$ 
which is lower than the reasonable value for the case of RNe. 

   As is mentioned above, the predicted outburst of T Pyx by the end of 1980th based on 
the calculated recurrence (about 20 years) was not detected by any observers in the world 
(or missing, but not plausible). 
After 45-year absence from outburst, the sixth outburst was detected by M. Linnolt (USA) 
at a visual magnitude of 13.0 on 2011 April 14.2931 UT (\cite{waa2011}).
The early stage light curve of the 2011 outburst was obtained by 
Solar Mass Ejection Imager (SMEI) observations (\cite{hou2011}). 
\citet{sho2011} derived an extinction ($E_{B-V} \sim 0.5 \pm 0.1$) from 
high resolution spectroscopy during 2011 outburst 
(total night of their observation was 7, among which 4 nights are at pre-maximum stage). 
\citet{che2011} performed near-IR photometric and spectroscopic 
observations on 2.37 day to 48.2 day after the outburst. 
They suggested an existence of face-on bipolar ejection from their analysis. 

   The main purpose of this paper is to present the detailed optical spectroscopic observational 
report of T Pyx during the early stage (pre-maximum to shortly after maximum) of its eruption. 
This report will give the comprehension of this RN's evolutional feature at its pre-maximum stage. 
In addition it is important to examine whether any spectral deference exists between 
the present outburst and past ones due to longer (45-year) quiescence.
In section 2, we present our spectroscopic observations. 
Section 3 gives the obtained results. 
Section 4 is for the discussions. Summary is stated in section 5.

\section{Observations}
  We have performed low-resolution spectroscopic observations from 2011 April 16 to May 14 
at the Tanabe Personal Observatory \footnote{
TPO's position is \timeform{34D42'42"}N, \timeform{133D52'21"}E. 
This observatory plays a complimentary role of Okayama University of Science (OUS) Observatory. 
The latter observatory is suitable for rising objects in the eastern sky and TPO is vise versa.}
(TPO) located at the northern part of Okayama-city. 
The total number of observation is 11 nights. 
Among them, 9 nights are at pre-maximum stage.
At the final stage of our observation, 
the angular distance between T Pyx and the sun became so small that
most of our observations were performed in the evening twilight. 
Table 1 is a journal of the spectroscopic observations. Our spectroscopic observational system 
is a combination of DSS-7 (SBIG production) spectrometer and ST-402 (SBIG) 
CCD (Kodak KAF-0402ME: $765 \times 510$ pixels) camera
installed on Celestron 28cm (F/10) Schmidt-Cassegrain telescope. 
The spectrometer's resolution $R = \lambda / \Delta\lambda$ is approximately 400 at 
6000 \AA, and its dispersion is 5.4 \AA $/$pixel. 
Hence the precision of the radial velocity is $\sim 700$ km $^{-1}$. 
Covering wavelength range is a 4000--8000 \AA. 
Wavelength calibrations is made by Hydrogen-Helium lamp.

\begin{table}
  \caption{Journal of spectroscopic observations. $V$ magnitude data from VSOLJ.}
  \begin{center}
 \begin{tabular}{lccc} \hline
Date (2011) &JD\footnotemark[$\dagger$]& $\Delta t$ (day)\footnotemark[$\ddagger$]& $V$ mag \\ \hline
Apr 16      & 5668.0                   & --25.6                                   & 7.9     \\
Apr 19      & 5671.0                   & --22.6                                   & 7.7     \\
Apr 20      & 5672.0                   & --21.6                                   & 7.6     \\
Apr 23      & 5675.0                   & --18.6                                   & 7.7     \\
Apr 24      & 5675.9                   & --17.7                                   & 7.7     \\
Apr 25      & 5676.9                   & --16.7                                   & 7.6     \\
Apr 29      & 5680.9                   & --12.7                                   & 7.1     \\
May 5       & 5687.0                   & --6.6                                    & 6.8     \\
May 8       & 5689.9                   & --3.7                                    & 7.0     \\
May 13      & 5695.0                   &  +1.4                                    & 6.6     \\
May 14      & 5696.0                   &  +2.4                                    & 6.6     \\
 \hline
 \end{tabular}
 \end{center}
 \footnotesize {\bf Note.}
 \noindent 
 \footnotemark[$\dagger$] JD$-2450000$\\
 \footnotemark[$\ddagger$] Days from maximum light (max date $= 2455693.6 $ JD at $6.3V$), obtained from the data collected by VSOLJ.
\end{table}

\section{Results}
  
  Figure 1 shows representative spectra of T Pyx at its early stage of 2011 outburst
\footnote{All of the spectra are available as on-line material.\\ {\tt http://...}}. 
On $\Delta t = -25.6$ to $-21.6$ days (where $\Delta t$ denotes the elapsed days from maximum light, 
which is determined by VSOLJ's photometric data, see upper panel of Fig. 3), 
we can see such broad and prominent emission lines as Balmer series, He I (5016, 5876, 6678), He II (4686), 
N II (4489, 5001, 5679, 5938, 6482), N III (4640) and O I (7773) in the upper panel (a) of figure 1.
Here the Balmer lines show clear P Cygni profile. 
Maximum expansion velocity derived from this profile is approximately 2200 km s$^{-1}$.
Helium and Nitrogen lines are weakened day by day.
According to the classification by \citet{wil1992}, 
such novae with these terms are to be classified as a He/N type. 

  On $\Delta t = -18.6$ days, however, Fe II (multiplets; 27, 28, 37, 38, 42, 48, 49 and 74) 
lines emerge on the spectra which can be seen in the lower panel (b) of figure 1. 
Fe II lines are strengthened day by day, and the P Cygni profiles also become more prominent than before. 
According to the classification by \citet{wil1992}, such novae with these terms are to be classified as a Fe II type. 
It is remarkable that the spectral type of T Pyx during the pre-maximum stage 
evolves from He/N type to Fe II type
\footnote{\citet{wil2012} calls such a type of spectral change ``hybrid evolution''.}. 

  Figure 2 is an enlarged spectra of H$\alpha$, showing the temporal behavior of its P Cygni profile. 
The abscissa is not wavelength but the radial velocity relative to the peak of H$\alpha$'s emission. 
The ordinate is normalized value of the spectral intensity. 
At first (He/N phase), H$\alpha$ shows clear P Cygni profile. 
However, this profile is temporarily diminished at earlier Fe II phase 
($\Delta t= -18.6$ to $-12.7$ days). 
Subsequently, it become clear again at later Fe II phase (after $\Delta t = -6.6$ days).
In addition, the intensity of absorption component is stronger at later Fe II phase than He/N one. 
These behavior may be explained by the increase of the optical thickness.

  Upper panel of figure 3 shows multi-color light curves ($B, V, R_c, I_c$) collected by VSOLJ. 
Temporal variations of the expansion velocity due to P Cygni profile 
(H$\alpha$, $\beta$, O I and Fe II) are shown in the lower panel. 
On $\Delta t = -25.6$ days, the blueshifted absorption minima of these lines 
show the expansion velocities of approximately 1600 km s$^{-1}$ (O I), 1900 km s$^{-1}$ (H$\beta$) 
and 2200 km s$^{-1}$ (H$\alpha$). 
It is also notable that after rapid decrease the expansion velocity shows gradual increase.

    \begin{figure*}
      \begin{center}
        \FigureFile(180mm,180mm){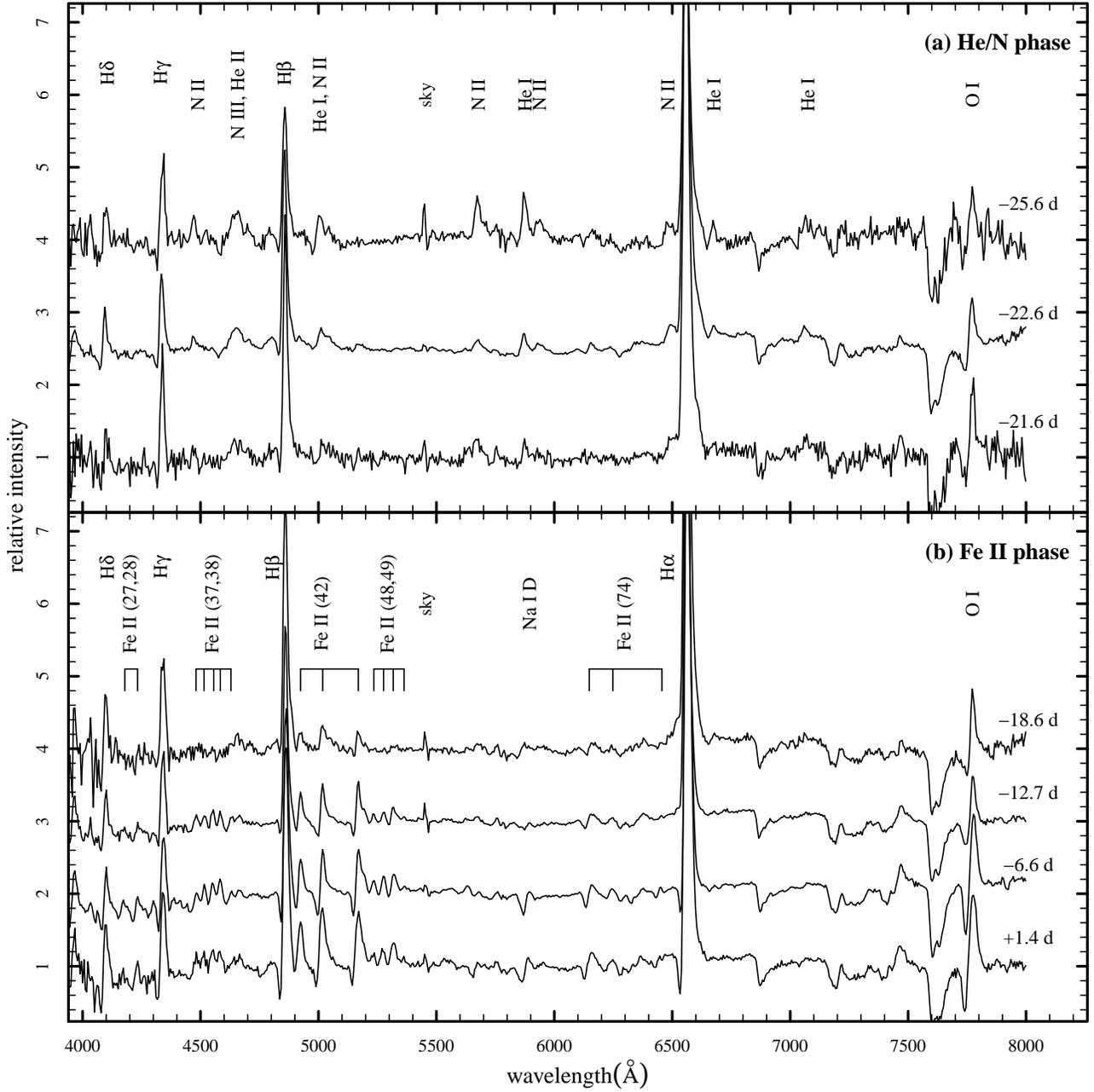}
      \end{center}
     \caption{Representative spectra of T Pyx from the pre-maximum to early post maximum stage. 
              The resolution power $R = \lambda / \Delta\lambda$ is $\approx400$ at 6000 \AA.
              The numerical values on the right edge are the elapsed days ($\Delta t$) from the maximum light.
              Each of these continuum is normalized as unity.
              An upper (a) and lower (b) panel show the earlier (He/N phase) spectra and 
              later (Fe II phase) ones, respectively.}
    \end{figure*}

\section{Discussion}

    \begin{figure}
      \begin{center}
        \FigureFile(80mm,80mm){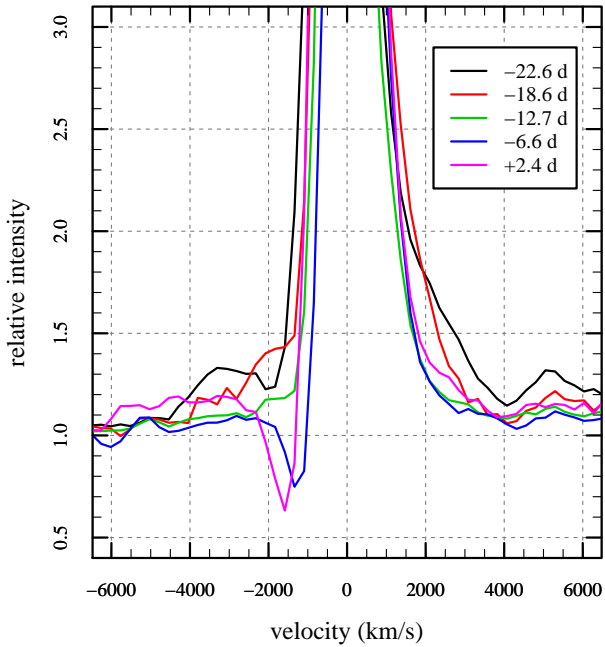}
      \end{center}
     \caption{Temporal change of H$\alpha$ profiles. Each of these continuum is normalized as unity. 
              The abscissa is radial velocity relative to the peak of H$\alpha$. 
              Black line is He/N phase. Red, green, blue and pink lines are Fe II one.}
    \end{figure}

    \begin{figure}
      \begin{center}
        \FigureFile(80mm,80mm){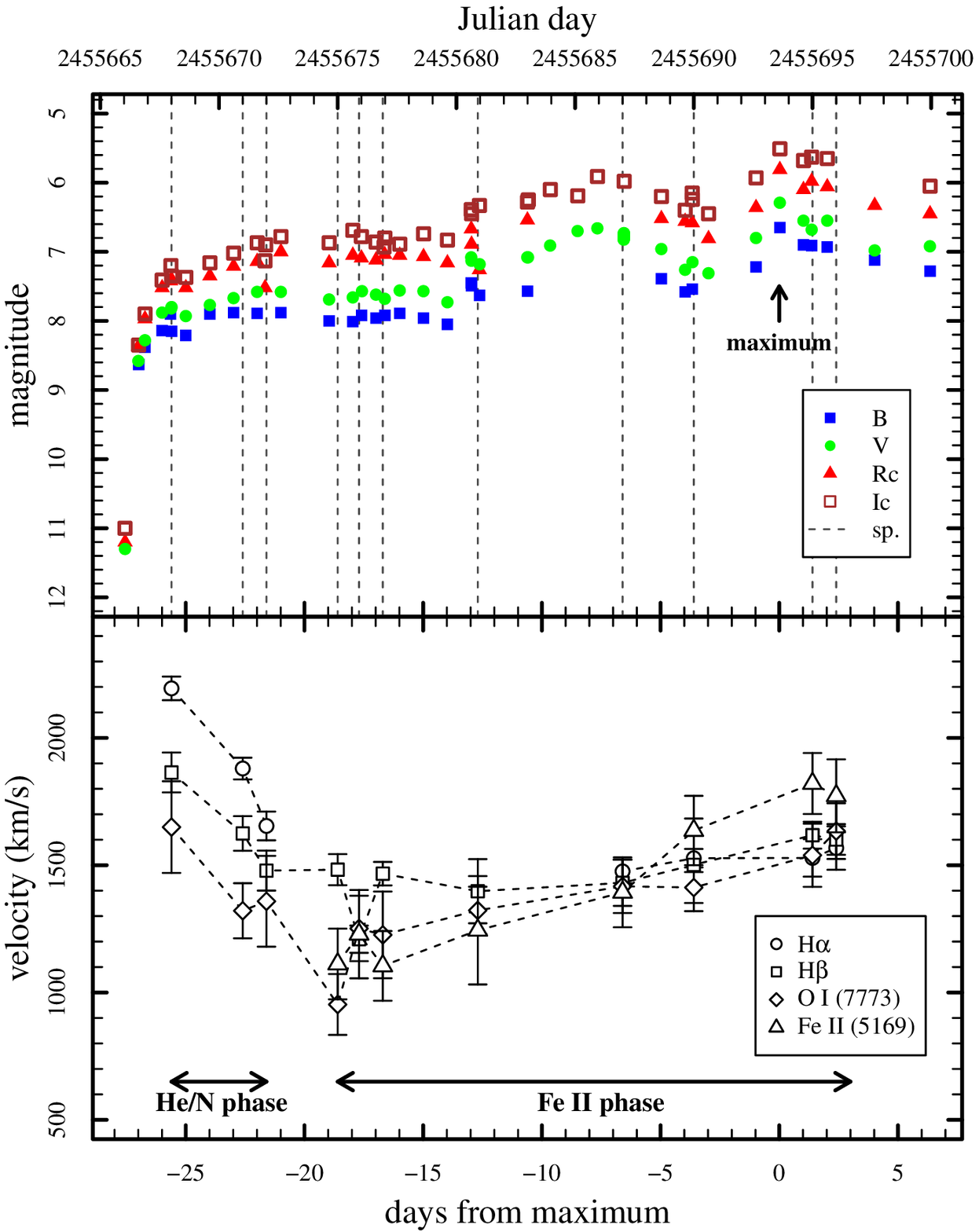}
      \end{center}
     \caption{{\bf Upper panel:} Light curves of T Pyx 2011 outburst during the pre-maximum and 
              maximum stages. Multi-color photometric data from VSOLJ. 
              Dashed lines (sp.) indicate the corresponding epochs to 
              the date of our spectroscopy. According to the light curves, maximum date is 
              thought to be May 12 (2455693.6 JD) at $6.3 V$. \\
              {\bf Lower panel:} Temporal change of the expansion velocity calculated by
               P Cygni profile (H$\alpha$, $\beta$, O I and Fe II). The precision of the 
               radial velocity from P Cygni profile is less than $700$ km s$^{-1}$.}
    \end{figure}

\subsection{The spectral type and its evolution}
  According to \citet{wil1992}, those novae that show stronger lines of He and N have larger expansion 
velocities and higher level of ionization. On the other hand, novae with prominent Fe II 
lines have slower expansion velocities and lower level of ionization, accompanying P Cygni absorption 
components. Hence the spectral type of T Pyx during the pre-maximum stage is thought to evolve from 
He/N type to Fe II type. 

  The pre-maximum spectra have generally been explained by a uniformly expanding optically thick envelope, 
cooling as it expand (Warner 1995, 2008). 
However, such a picture is difficult for explaining the behavior of T Pyx at its pre-maximum spectral evolution. 
In this case, the following sequential process is thought to be occurred: 
 \def\theenumi{\roman{enumi}}
 \begin{enumerate}
 \item At the earlier pre-maximum stage, an optically thin region is formed 
       as a result of rapid expansion by nova eruption, resulting in He/N type spectrum. 
 \item At the later stage an optically thick region is formed due to suppression effect by 
       growing photosphere, resulting in Fe II type spectrum. 
\end{enumerate}
In accordance with these two steps, the expansion velocity changes from decreasing to increasing mode 
(see lower panel of Fig. 3).

\subsection{Comparison with the previous outburst}
  \citet{cat1969} had obtained spectra of T Pyx 1966 outburst 
on 23 and 20 days before maximum light (unfortunately only two nights). 
The emission lines of Balmer series, He I, N II, O II, Mg II were seen on $\Delta t = -23$ days. 
He reported that N II line became weaken and Fe II lines emerged on $\Delta t = -20$ days. 
From our results, N II lines can be seen on $\Delta t= -25.6$ to $-21.6$ 
days, and then Fe II lines can be seen from $\Delta t = -18.6$ days. 
The spectral evolution of T Pyx in 2011 outburst seems to be consistent with 1966 outburst. 

\subsection{Comparison with other novae}
  So far, the only known nova showing the pre-maximum spectral type evolution  
from He/N to Fe II type is a classical nova (CN) V5558 Sgr (\cite{tan2011}). 
However this CN shows much slower ($\sim 500$ km s$^{-1}$) and monotonously increasing expansion velocity. 
Concerning the velocity variation at pre-maximum stage, CN V723 Cas is similar to T Pyx. 
As is reported by \citet{iij1998} this CN's velocity behavior at pre-maximum stage is from 
decreasing phase to increasing one. However this CN seems to show no spectral evolution during 
the pre-maximum stage.

\section{Summary}

\def\theenumi{\Roman{enumi}}
\begin{enumerate}
 \item The spectral type of T Pyx during the pre-maximum stage is thought to evolve from 
       He/N type to Fe II one.  
 \item The velocity from blueshifted absorption component is at first in decreasing mode, 
       then turns to be increasing mode.
 \item Spectral evolution of T Pyx 2011 outburst during pre-maximum stage is thought to be 
       compatible with 1966 outburst. 
 \item The spectral evolution from He/N type to Fe II type during the pre-maximum stage is similar to 
       V5558 Sgr, but temporal change of expansion velocity is similar to V723 Cas rather than V5558 Sgr.
\end{enumerate}

 \vspace{10pt}
 We would like to express gratitude to VSOLJ (S. Kiyota, H. Maehara and H. Itoh) 
 for their useful photometric data.

\end{document}